\documentclass[aps,prd,reprint,superscriptaddress]{revtex4-2}
\usepackage{amsmath, amssymb, bm, ulem, xcolor}
\usepackage{graphicx}

\begin{document}

\title{\boldmath Hadroproduction data support tetraquark hypothesis for
$\chi_{c1} (3872)$}

\author{Wai Kin Lai}
\email{samlai2357@gmail.com}
\affiliation{State Key Laboratory of Nuclear Physics and Technology, Institute of
Quantum Matter, South China Normal University, Guangzhou 510006, China}
\affiliation{Guangdong Basic Research Center of Excellence for Structure and Fundamental Interactions of Matter,
Guangdong Provincial Key Laboratory of Nuclear Science, Guangzhou 510006, China}

\author{Hee Sok Chung}
\email{heesokchung@gwnu.ac.kr}
\affiliation{Department of Mathematics and Physics,
Gangneung-Wonju National University, Gangneung 25457, Korea}

\date{\today}

\begin{abstract}
We show that the recently proposed tetraquark hypothesis for the nature of the
$\chi_{c1}(3872)$ results in a formalism for inclusive production rates that
has no unknown parameters. We employ this formalism to compute hadroproduction
rates of $\chi_{c1}(3872)$ at the Large Hadron Collider, which agree with
measured prompt and nonprompt cross sections. 
Thus, we find that the tetraquark hypothesis for $\chi_{c1}(3872)$ is 
well supported by hadroproduction data. 
\end{abstract}

\maketitle

\section{Introduction}
The discovery of the narrow charmoniumlike state with a
mass of about 3.872~GeV~\cite{Belle:2003nnu,
CDF:2003cab,D0:2004zmu,BaBar:2004oro}, now known as the $\chi_{c1}(3872)$, led
to a renewed interest in exotic hadrons since their existence were prophesized
in the 1960s~\cite{Gell-Mann:1964ewy,Zweig:1964ruk}.  While this state carries
quantum numbers that are consistent with a conventional
charmonium~\cite{LHCb:2013kgk, LHCb:2015jfc}, it is widely considered a
candidate for an exotic hadron~\cite{Barnes:2003vb}.  There have been a number
of hypotheses for its nature including it being a $D^{*0} \bar{D}^0$
molecule~\cite{Tornqvist:2004qy,Voloshin:2003nt,Close:2003sg,Wong:2003xk,
Braaten:2003he,Swanson:2003tb},
a hybrid~\cite{Close:2003mb,Li:2004sta}, a tetraquark~\cite{Maiani:2004vq}, 
and even a conventional
charmonium~\cite{Eichten:2002qv,Eichten:2004uh}, but there seems to
be no consensus on which the correct interpretation of this state is. 

Inclusive production rates of $\chi_{c1}(3872)$ at hadron colliders has
been considered a test ground for these hypotheses~\cite{Bignamini:2009sk}. 
The prompt and nonprompt cross
sections as functions of the transverse momentum $p_T$ have been measured at the
Large Hadron Collider (LHC) by the CMS~\cite{CMS:2013fpt},
ATLAS~\cite{ATLAS:2016kwu}, and LHCb~\cite{LHCb:2021ten} Collaborations. 
The prompt measurements are, for example, in conflict with the
molecular~\cite{Artoisenet:2009wk} and the
charmonium~\cite{Butenschoen:2013pxa} hypotheses; 
while the mixing model in Refs.~\cite{Meng:2005er, Meng:2013gga} can explain
the prompt rates, this comes at the cost of two nonperturbative unknowns that
cannot be computed from first principles and must be determined from data.

Recently, in Refs.~\cite{Grinstein:2024rcu, Braaten:2024tbm, BMSV}, the authors 
proposed a scenario based on the Born-Oppenheimer approximation of QCD 
where the $\chi_{c1}(3872)$ can arise as a $c \bar c q \bar q$ state, where the
$c \bar c$ is predominantly in a color-octet state at short distances. 
While these hypotheses can be formulated in the Born-Oppenheimer effective field theory (BOEFT)
approach~\cite{Born:1927rpw, Juge:1999ie, Braaten:2014qka,Berwein:2015vca,Oncala:2017hop,
Brambilla:2018pyn,Brambilla:2019jfi, Maiani:2019cwl,Berwein:2024ztx},
because
some of the potentials used in Refs.~\cite{Grinstein:2024rcu, Braaten:2024tbm,  
BMSV} have not yet been obtained from
first principles and instead are tuned to the spectrum, we may not yet consider
this hypothesis a solid prediction of QCD. This calls for independent tests to
be made to compare this scenario with measurements. 

In this paper, we compute hadroproduction cross sections of the
$\chi_{c1}(3872)$ under the assumption that it is a $c \bar c q \bar q$
tetraquark state as proposed in Refs.~\cite{Grinstein:2024rcu, 
Braaten:2024tbm,  BMSV}. 
We find that, due to the color and angular momentum configuration of the 
$c \bar c$ at short distances in this hypothesis, the inclusive production rate
takes a particularly simple form in the nonrelativistic QCD (NRQCD)
factorization formalism~\cite{Bodwin:1994jh}, because the $\chi_{c1}(3872)$ is
predominantly produced via a single color-octet state. 
Moreover, the nonperturbative NRQCD matrix element, which
corresponds to the probability to find the $c \bar c$ inside the
$\chi_{c1}(3872)$ in the color-octet state, can be computed from the $c \bar c$
wavefunction, which, in turn, can be determined from a Schr\"odinger equation.
This leads to a formalism for inclusive production of $\chi_{c1}(3872)$ with no
unknown parameters, which can be used as a stringent test of the tetraquark
hypothesis for $\chi_{c1}(3872)$.  We compare our calculation of prompt and
nonprompt hadroproduction rates of $\chi_{c1}(3872)$ from $pp$ collisions at
the LHC and compare them with available data. 

\section{\boldmath NRQCD factorization for $\chi_{c1}(3872)$}
In the NRQCD factorization 
formalism~\cite{Bodwin:1994jh}, the inclusive cross section of a
$\chi_{c1}(3872)$ can be written as sums of products of 
the cross section of a $c \bar c$ in a specific color and angular momentum
state and the nonperturbative NRQCD matrix element that corresponds to the 
probability to find the $c \bar c$ inside the $\chi_{c1}(3872)$ in the
corresponding color and angular momentum state. 
The most important matrix element is the lowest-dimensional one that
corresponds to the dominant $c \bar c$ Fock state of the $\chi_{c1}(3872)$. 
The contributions from matrix elements of higher dimensions, as well as matrix
elements corresponding to subleading Fock states, are
suppressed by powers of the scale $\Lambda$ of the BOEFT divided by the charm
quark mass $m_c$. 
In the case of the $\chi_{c1}(3872)$ in the tetraquark scenario, 
the dominant $c \bar{c}$ Fock state is the 
color-octet spin-triplet $S$-wave ($^3S_1^{[8]}$)
state~\cite{Grinstein:2024rcu, Braaten:2024tbm, BMSV}, which corresponds to a matrix element of 
the lowest possible dimension (dimension 3). Hence, we have 
\begin{equation}
\label{eq:NRQCDfac}
\sigma_{\chi_{c1}(3872)} = 
\sigma_{c \bar c (^3S_1^{[8]})} 
\langle {\cal O}^{\chi_{c1}(3872)}(^3S_1^{[8]}) \rangle, 
\end{equation}
where $\sigma_{c \bar c (^3S_1^{[8]})}$ is the perturbatively calculable
production rate of a $c \bar c$ in the $^3S_1^{[8]}$ state 
and $\langle {\cal O}^{\chi_{c1}(3872)}(^3S_1^{[8]}) \rangle$ 
is the corresponding NRQCD matrix element. 
A striking feature of this formula is that only one channel contributes to the
cross section at leading power in the nonrelativistic expansion. 
This is because the matrix elements that
correspond to Fock states other than the $^3S_1^{[8]}$ one are
always of higher dimensions; for example, the color-octet $D$-wave component
contributes through a dimension-7 matrix element, which is 
suppressed by at least $(\Lambda/m_c)^4$ compared to the $^3S_1^{[8]}$
contribution. The color-singlet $P$-wave component contributes through
dimension-5 matrix elements, which are suppressed by at least $(\Lambda/m_c)^2$
compared to the leading contribution; this is further suppressed due to the
fact that the regular charmonium component is small in the $\chi_{c1}(3872)$
state~\cite{BMSV}. 
This is in stark contrast with the case of $J/\psi$ or $\psi(2S)$ production
phenomenology, where, although only color-singlet matrix elements appear 
at leading power in the nonrelativistic expansion, it is necessary to include 
contributions from color-octet matrix elements with suppressed power counts 
in the factorization formula, because production rates of a color-octet 
$c \bar c$ are much greater than that of a color singlet and can dominate the
cross section. 
This does not happen in the case of $\chi_{c1}(3872)$, because the $^3S_1^{[8]}$
channel already has a large $c \bar c$ production rate and there is no source
of dynamical enhancement from power-suppressed channels. 

\section{Prediction of nonprompt fraction}
We can already make predictions based
on Eq.~(\ref{eq:NRQCDfac}) by using the fact that, since the formula involves
only one channel, the NRQCD matrix element cancels in cross section ratios. 
The nonprompt fraction, which is the ratio of the
nonprompt cross section to the 
sum of the prompt and nonprompt cross sections, has been
measured by CMS~\cite{CMS:2013fpt} and ATLAS~\cite{ATLAS:2016kwu}. 
We can compute this fraction by using Eq.~(\ref{eq:NRQCDfac}) as
\begin{equation}
\label{eq:nonpromptfrac}
\frac{\sigma^{\rm nonprompt}_{\chi_{c1}(3872)}}
{\sigma^{\rm prompt}_{\chi_{c1}(3872)} + 
\sigma^{\rm nonprompt}_{\chi_{c1}(3872)}}
= 
\frac{\sigma^{\rm nonprompt}_{c \bar c (^3S_1^{[8]})}}
{\sigma^{\rm prompt}_{c \bar c (^3S_1^{[8]})} + 
\sigma^{\rm nonprompt}_{c \bar c (^3S_1^{[8]})}}. 
\end{equation}
Here, $\sigma^{\rm nonprompt}_{c \bar c (^3S_1^{[8]})}$ is given by 
\begin{equation}
\label{eq:nonpromptrate}
\sigma^{\rm nonprompt}_{c \bar c (^3S_1^{[8]})}
= 2 \times \sigma_b \times C(b \to c \bar c (^3S_1^{[8]}) +X), 
\end{equation}
where the factor $2$ counts for the fact that the nonprompt cross section comes
from both $b$ and $\bar b$ decays, $\sigma_b$ is the production rate of a
$b$ quark, and $C(b \to c \bar c (^3S_1^{[8]}) +X)$ is the NRQCD
short-distance coefficient for the branching fraction 
${\rm Br}(b \to \chi_{c1}(3872)+X) = C(b \to c \bar c (^3S_1^{[8]}) +X) 
\times \langle {\cal O}^{\chi_{c1}(3872)}(^3S_1^{[8]}) \rangle$.
This short-distance coefficient has been computed at next-to-leading order 
(NLO) in Ref.~\cite{Beneke:1998ks}, which gives 
$C(b \to c \bar c (^3S_1^{[8]}) +X) = 0.223$~GeV${}^{-3}$ 
in the 't\,Hooft--Veltman scheme. 
We take the fixed order+next-to-leading log (FONLL) result for 
$\sigma_b$~\cite{Cacciari:1998it, Cacciari:2001td}, in which we
include the correction from the nonperturbative fragmentation function.  
We compute the prompt cross section in two ways. In fixed-order (FO) 
perturbation theory, we compute the prompt $c \bar c (^3S_1^{[8]})$ cross
section at NLO by using the {\sc fdchqhp} package~\cite{Wan:2014vka}. Because the prompt cross
section is dominated by the fragmentation process which involves singularities
at the kinematical threshold, threshold logarithms that appear in radiative
corrections can become significant; therefore, we also compute the 
{\it resummed} prompt cross section by using the results of 
Ref.~\cite{Chung:2024jfk}, 
where the threshold double logarithms, as well as the
Dokshitzer-Gribov-Lipatov-Altarelli-Parisi (DGLAP)
logarithms~\cite{Gribov:1972ri, Lipatov:1974qm, Dokshitzer:1977sg,
Altarelli:1977zs}, are resummed to all orders in perturbation theory. We note
that the DGLAP logarithms are also resummed in the FONLL calculation of
$\sigma_b$. The numerical effect of threshold resummation in the prompt cross
section is similar to the inclusion of the nonperturbative fragmentation
function in FONLL, as both have the effect of replacing
the singular distribution in
the fragmentation function with a smooth one. Neither resummation of threshold
or DGLAP logarithms are included in the FO calculation. We choose $m_c =
1.5$~GeV, which is the default choice in Refs.~\cite{Cacciari:1998it,
Cacciari:2001td, Wan:2014vka}, and set the scales to $p_T$.

\begin{figure}[t]
\includegraphics[width=\columnwidth]{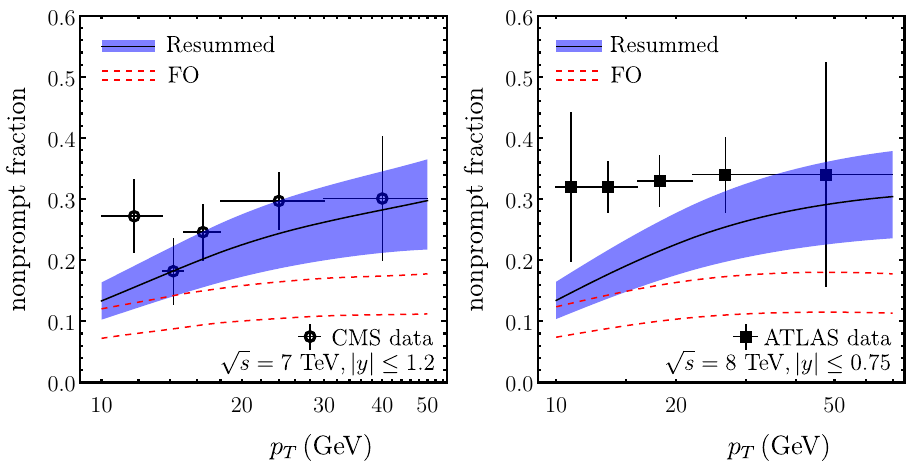}%
\caption{\label{fig:fraction} 
Nonprompt fractions for $\chi_{c1}(3872)$ production computed from
Eq.~(\ref{eq:nonpromptfrac}) compared to CMS (left) and ATLAS (right) data. 
}
\end{figure}

We compare our  results for the nonprompt fraction with CMS and ATLAS
data in Fig.~\ref{fig:fraction}. We include uncertainties from varying the
scales by factors $1/2$ and $2$, the uncertainty from the $b$ quark mass in the
FONLL calculation, and estimate the uncertainty in the theoretical calculation
of the branching fraction from the dependence on the $\gamma_5$ scheme by
the difference between the results in the 't\,Hooft--Veltman and the na\"ive
dimensional regularization schemes, which amounts to about $22\%$. 
The uncertainties are added in quadrature. We see that, when we
include resummation when computing both the prompt and nonprompt cross
sections, our predictions are in fair agreement with the CMS data 
and also with ATLAS data at large $p_T$. 
The FO results, where effects of resummation are neglected in the prompt
contribution, are systematically below data. 
The fact that our prediction for the nonprompt fraction that 
requires no unknown nonperturbative parameters agree with measurements 
makes a very compelling case for the tetraquark hypothesis for 
$\chi_{c1}(3872)$.

\section{Prompt and nonprompt cross sections}
\subsection{\boldmath NRQCD matrix element for $\chi_{c1}(3872)$}
Another important test of
the scenario in Refs.~\cite{Grinstein:2024rcu, Braaten:2024tbm, BMSV} is to see if we can reproduce the absolute cross sections
at hadron colliders, which require knowledge of the $^3S_1^{[8]}$ matrix
element. By using the techniques developed in Refs.~\cite{Brambilla:2020ojz,
Brambilla:2021abf, Brambilla:2022rjd, Brambilla:2022ayc}, we compute this
matrix element in terms of the $c \bar c$ wavefunction of the
$\chi_{c1}(3872)$.  The calculation is almost identical to the case of the
$^3S_1^{[1]}$ matrix element for $J/\psi$, except that for $\chi_{c1}(3872)$,
the $c \bar c$ is in a color-octet state.  The $^3S_1^{[8]}$ matrix element is
defined by the vacuum expectation value of the
operator~\cite{Bodwin:1994jh,Nayak:2005rw} 
\begin{align}
\label{eq:3S18op}
\chi^\dag \sigma^i T^a \psi \Phi^{\dag ab}_\ell
{\cal P}_{\chi_{c1}(3872)}
\Phi^{bc}_\ell \psi^\dag \sigma^i T^c \chi,
\end{align}
where $\psi$ and $\chi$ are Pauli spinor fields that annihilate and create a
charm quark and antiquark, respectively, $\sigma^i$ are Pauli matrices, 
$T^a$ are color matrices, and $\Phi_{\ell} = {\cal P} \exp[-i g \int_0^\infty d
\lambda \ell \cdot A^{\rm adj} (\lambda \ell)]$ is a path-ordered Wilson line 
in the adjoint representation defined along a lightlike direction $\ell$. The
${\cal P}_{\chi_{c1}(3872)}$ is an operator that projects onto states that
include a $\chi_{c1}(3872)$ at rest. 
This operator can be written as 
\begin{align}
& {\cal P}_{\chi_{c1}(3872)}
= 
\int d^{d-1}x_1 d^{d-1}x_2 d^{d-1}x_1' d^{d-1}x_2' 
\nonumber \\ & 
\sum_{n \in {\mathbb O}} 
\phi(\bm{x}_1'-\bm{x}_2') 
| \underline{\rm n};\bm{x}_1',\bm{x}_2' \rangle
\langle \underline{\rm n};\bm{x}_1,\bm{x}_2 | 
\phi^* (\bm{x}_1-\bm{x}_2), 
\end{align}
where $d$ is the number of spacetime dimensions,
$\phi(\bm{x}_1-\bm{x}_2)$ is the unit-normalized $c \bar c$ wavefunction 
with $\bm{x}_1$ and $\bm{x}_2$ the position of the $c$ and $\bar c$, respectively, and $|\underline{\rm n};\bm{x}_1,\bm{x}_2 \rangle$ is the NRQCD state
of an excitation we denote with $n$.
Here, we used the heavy-quark spin symmetry to lift the correlation between the
angular momentum of the $c \bar c$ and the light degrees of freedom. 
The sum over $n$ is restricted to the subspace
${\mathbb O}$ where the $c \bar c$ is in the color-octet state in the limit
$\bm{x}_1 -\bm{x}_2 \to 0$. To compute the vacuum expectation value of
Eq.~(\ref{eq:3S18op}), we need the matrix element of $\Phi^{bc}_\ell \psi^\dag
\sigma^i T^c \chi$ between the QCD vacuum and the state $\langle \underline{\rm
n};\bm{x}_1,\bm{x}_2 |$. By using the techniques used in
Refs.~\cite{Brambilla:2020ojz, Brambilla:2021abf, Brambilla:2022rjd,
Brambilla:2022ayc}, we obtain at leading power in the nonrelativistic expansion
\begin{align}
& \langle \underline{\rm n};\bm{x}_1,\bm{x}_2 |
\Phi^{bc}_\ell \psi^\dag \bm{\sigma} T^c \chi
| 0 \rangle
= 
\delta^{(d-1)} (\bm{x}-\bm{x}_1) 
\nonumber \\ & \times
\delta^{(d-1)} (\bm{x}-\bm{x}_2)
\langle n; \bm{x}_1,\bm{x}_2 | \Phi^{bc}_\ell 
\bm{\sigma} T^d \Phi^{dc}_0 | 0 \rangle, 
\end{align}
where $|0\rangle$ is the QCD vacuum, 
$\bm{x}$ is the spatial position of the operator on the left side, 
$|\underline{\rm n}; \bm{x}_1,\bm{x}_2 \rangle
= \psi^\dag (\bm{x}_1) \chi(\bm{x}_2) | n; \bm{x}_1,\bm{x}_2 \rangle$, 
the spin and fundamental color indices of $\bm{\sigma}$ and $T^d$ are fully 
contracted with the ones implicit in the state $\langle n; \bm{x}_1,\bm{x}_2
|$, 
and $\Phi_0 = {\cal P} \exp[-i g \int_0^\infty d
\lambda A_0^{\rm adj} (\lambda,\bm{0})]$
is the temporal Wilson line in the adjoint representation. 
Because of this, this matrix element vanishes unless the $c \bar c$ is in
the color-octet state at $\bm{x}_1 - \bm{x}_2 \to \bm{0}$. Therefore, we can
lift the restriction $n \in {\mathbb O}$ in the sum and use the 
completeness relation 
$\sum_n | n; \bm{x}_1,\bm{x}_2 \rangle \langle n; \bm{x}_1,\bm{x}_2 | = 1$, 
which leads to 
\begin{align}
\langle {\cal O}^{\chi_{c1}(3872)}(^3S_1^{[8]}) \rangle
= 2 T_F (d-1) {\cal S} | \phi(\bm{0}) |^2 , 
\end{align}
where the factors $2 (d-1)$ and $T_F=1/2$ come from the trace over spin and color,
respectively, 
$\phi(\bm{0})$ is the $\chi_{c1} (3872)$ wavefunction at the origin,
and ${\cal S}$ is a dimensionless quantity defined by 
the vacuum expectation value of adjoint Wilson lines:
\begin{align}
\label{eq:soft3S1}
{\cal S} &=
\langle 0 | [ \Phi_0^{ca} \Phi_\ell^{ba} ]^\dag 
\Phi_0^{cd} \Phi_\ell^{bd} | 0 \rangle.
\end{align}
In perturbation theory, ${\cal S} = 1$ at tree level, and the NLO correction
vanishes~\cite{Chung:2024jfk}. As we work at NLO accuracy, 
we take the perturbative QCD result ${\cal S} = 1 + O(\alpha_s^2)$.  
We then obtain 
\begin{align}
\label{eq:3s18me}
\langle {\cal O}^{\chi_{c1}(3872)}(^3S_1^{[8]}) \rangle = 3 | \phi(\bm{0}) |^2.
\end{align}
Because we can compute the wavefunction $\phi$ by solving a Schr\"odinger
equation, a potential model for the tetraquark completely determines the value
of the $^3S_1^{[8]}$ matrix element at leading power in the nonrelativistic
expansion. Therefore, the proposed scenario in Refs.~\cite{Grinstein:2024rcu,
Braaten:2024tbm, BMSV} leads to a formalism for inclusive production rates of $\chi_{c1}(3872)$
that has no unknown parameters. An analogous formalism would hold, for example,
for production of hybrid states with a $c \bar c$ component in the $S$-wave
state. 

\subsection{Prediction of cross sections}
In order to compute absolute cross
sections we must determine the $^3S_1^{[8]}$ matrix element from 
$|\phi(\bm{0})|^2$, which can be obtained by solving the Schr\"odinger
equation. In practice, however, we find that the result varies depending on 
the choice of potentials and parameters of the Schr\"odinger equation; 
we obtain values of $|\phi(\bm{0})|^2$ that range from $0.69 \times 10^{-3}$ to
$2.0 \times 10^{-3}$~GeV$^3$ from Refs.~\cite{Grinstein:2024rcu, Braaten:2024tbm, BMSV}. 
As our knowledge of the tetraquark potential is currently limited, for
obtaining 
phenomenological results in this work we choose to fix $|\phi(\bm{0})|^2$
against the experimental value ${\rm Br} (b\to \chi_{c1} (3872)+X) \times
{\rm Br} (\chi_{c1}(3872) \to J/\psi \pi^+ \pi^-) = 
(4.3 \pm 0.5) \times 10^{-5}$ from
LHCb~\cite{LHCb:2021ten} and the Particle Data Group value ${\rm Br} (\chi_{c1}
(3872) \to J/\psi \pi^+ \pi^-) = (3.5 \pm 0.9) \times
10^{-2}$~\cite{ParticleDataGroup:2024}. 
This corresponds to the central values $|\phi(\bm{0})|^2=1.8 \times
10^{-3}$~GeV$^3$ and $\langle {\cal O}^{\chi_{c1}(3872)}(^3S_1^{[8]}) \rangle =
5.5 \times 10^{-3}$~GeV${}^{-3}$, which lie within the range of potential 
model results. 
We consider the uncertainties in the $c \bar c$ cross sections by varying the
scales by factors $1/2$ and $2$. 
The uncertainty in $|\phi(\bm{0})|^2$ coming from the input parameters and the
theoretical uncertainty in  $C (b\to c \bar{c} (^3S_1^{[8]})+X)$ amount
to about 25\%. Because our result for 
$\langle {\cal O}^{\chi_{c1}(3872)}(^3S_1^{[8]}) \rangle$ can receive 
corrections of higher orders in the nonrelativistic expansion, we estimate the
total uncertainty in the matrix element to be 40\% of the central
value.

\begin{figure}[t]
\includegraphics[width=\columnwidth]{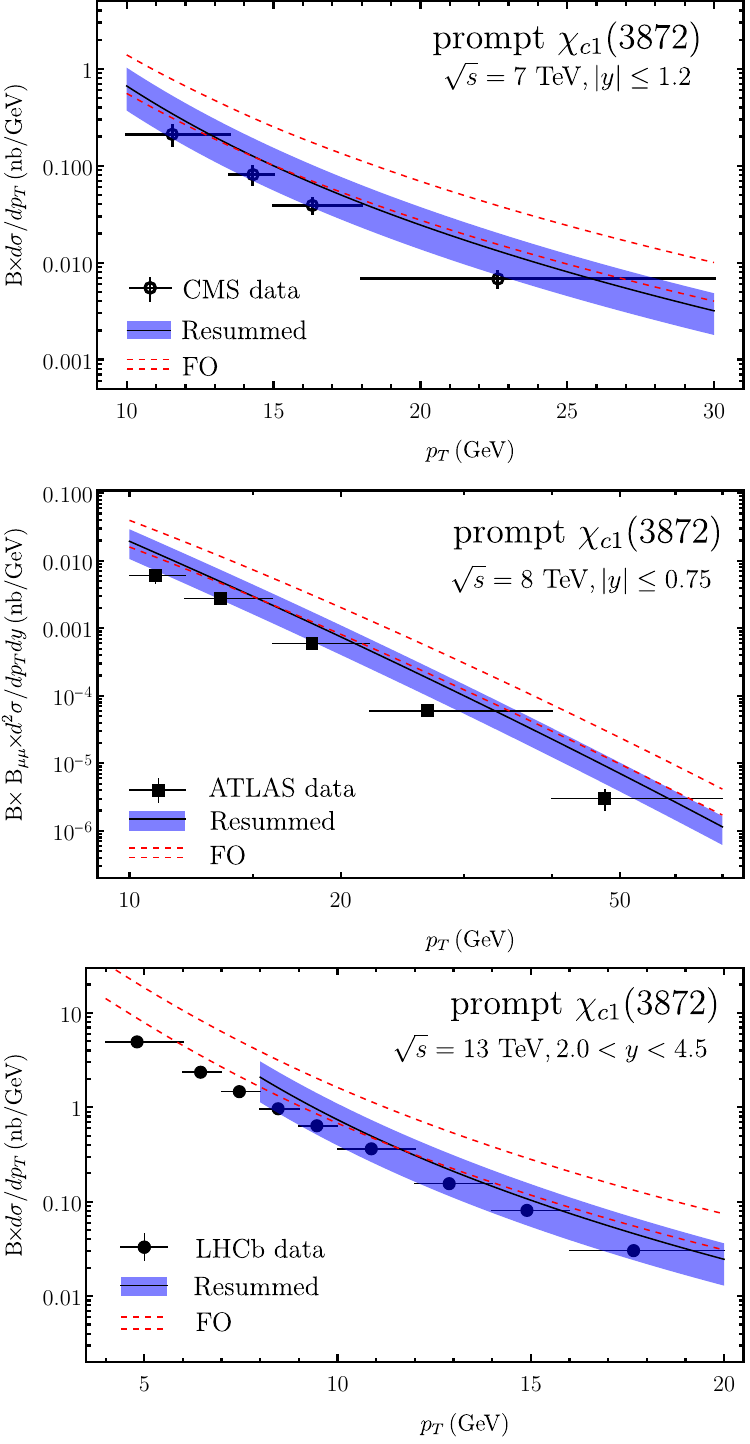}%
\caption{\label{fig:prompt} 
Prompt production rates of $\chi_{c1}(3872)$ computed in the tetraquark
hypothesis compared to CMS, ATLAS, and LHCb data. 
${\rm B} \equiv {\rm Br}(\chi_{c1}(3872) \to J/\psi \pi^+ \pi^-)$ 
and ${\rm B}_{\mu \mu} \equiv {\rm Br} (J/\psi \to \mu^+ \mu^-)$. 
}
\end{figure}

We first compute the prompt cross sections from $pp$ collisions at the LHC. 
The calculation of the $c \bar c$ cross sections are done in the same way as
in the nonprompt fractions. 
The comparison with CMS, ATLAS, and LHCb data are shown in
Fig.~\ref{fig:prompt}. 
The theoretical uncertainties come from scale variations and the 
$^3S_1^{[8]}$ matrix element. 
Our resummed results are in agreement with data, with the exception of ATLAS
data for the two largest $p_T$ bins. 
The FO results tend to overestimate data, although the FO and resummed results
are consistent within uncertainties. 
The difference between FO and resummed results diminish with
decreasing $p_T$. In the LHCb case, we do not show resummed
results below $p_T < 8$~GeV, because the cross section no longer becomes
dominated by fragmentation.

\begin{figure}[t]
\includegraphics[width=\columnwidth]{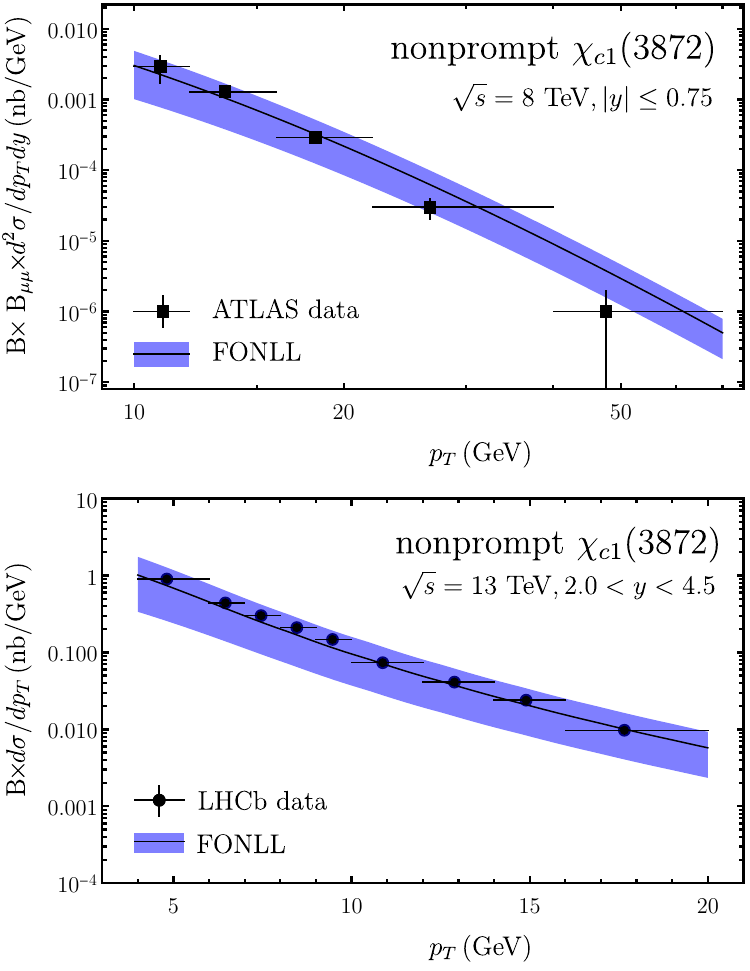}%
\caption{\label{fig:nonprompt} 
Nonprompt production rates of $\chi_{c1}(3872)$ computed in the tetraquark
hypothesis compared to ATLAS and LHCb data. 
${\rm B} \equiv {\rm Br}(\chi_{c1}(3872) \to J/\psi \pi^+ \pi^-)$ 
and ${\rm B}_{\mu \mu} \equiv {\rm Br} (J/\psi \to \mu^+ \mu^-)$. 
}
\end{figure}

We then compute the nonprompt cross sections from $pp$ collisions at the LHC. 
We again use the FONLL results for the $b$ production
rates~\cite{Cacciari:1998it, Cacciari:2001td, Wan:2014vka}. 
Our results compared with ATLAS and LHCb data are shown in 
Fig.~\ref{fig:nonprompt}. 
We see that our results are in good agreement with nonprompt measurements.

\subsection{Comparison with other model predictions}
As we have stated earlier, the
molecular~\cite{Artoisenet:2009wk} and conventional
charmonium~\cite{Butenschoen:2013pxa} hypotheses lead to predictions of the
prompt cross section that disagree with measurements~\cite{CMS:2013fpt}. 
In the mixing model~\cite{Meng:2005er, Meng:2013gga}, where the production rate
of the $\chi_{c1} (3872)$ is dominated by the conventional charmonium
component, cross section predictions require determination of two
nonperturbative unknowns: the $^3S_1^{[8]}$ matrix element of the $P$-wave
charmonium and $Z_{c \bar c}$, the amount of the conventional charmonium
admixture in the $\chi_{c1} (3872)$ state. In Ref.~\cite{Meng:2013gga}, these
unknowns have been obtained from fits to prompt cross section measurements.  
We find that they lead to values of 
${\rm Br} (b \to \chi_{c1} (3872)+X) {\rm Br}(\chi_{c1} (3872) \to J/\psi \pi^+
\pi^-)$ less than about $1.1 \times 10^{-5}$, which is more than 3 times
smaller than the LHCb result~\cite{LHCb:2021ten}. 
Hence, mixing-model predictions for nonprompt production rates and 
nonprompt fractions are generally below data.

\section{Conclusions}
In this work, we showed that the tetraquark hypothesis
for the nature of $\chi_{c1}(3872)$ proposed in Refs.~\cite{Grinstein:2024rcu, 
Braaten:2024tbm, BMSV} leads to a
formalism for its inclusive production rates that has no unknown parameters. 
Unlike the case of conventional quarkonium, the production rate of
$\chi_{c1}(3872)$ in this tetraquark hypothesis involves only one color-octet
channel in the nonrelativistic QCD factorization formalism, and the
nonperturbative matrix element can be determined from the $c \bar c$
wavefunction. We computed prompt and nonprompt production rates of
$\chi_{c1}(3872)$ from $pp$ collisions at the LHC, which are generally in
agreement with CMS, ATLAS, and LHCb measurements. This is a strong implication
that hadroproduction data support the tetraquark hypothesis for
$\chi_{c1}(3872)$. 

Currently, the precision of the theory prediction for absolute cross sections 
is limited by the lack of knowledge in the potentials of the BOEFT. While 
it seems promising that the nonperturbative matrix element fixed against
LHCb data~\cite{LHCb:2021ten} used in this work is within the range of
potential model calculations based on Refs.~\cite{Grinstein:2024rcu, Braaten:2024tbm, BMSV}, 
more theoretical progress will need to be made in order to reduce model
dependence. We also note that our theoretical results tend to overshoot data at
the largest $p_T$ bins, which may signal the need for resummation of threshold
logarithms beyond leading double logarithmic level~\cite{Chung:2024jfk}. 

If it turns out that the $\chi_{c1}(3872)$ is indeed a tetraquark as proposed 
in Refs.~\cite{Grinstein:2024rcu, Braaten:2024tbm, BMSV}, the production mechanism we established
in this work would open
new possibilities to probe QCD interactions through production of
$\chi_{c1}(3872)$ in colliders. Although the idea that a $c \bar c$ meson could
be produced predominantly through the $^3S_1^{[8]}$ channel is not
new~\cite{Braaten:1994vv}, 
it is now well known that the old idea of $^3S_1^{[8]}$ dominance for $J/\psi$ 
and $\psi(2S)$ production fails to describe large-$p_T$ hadroproduction data,
especially for polarized cross sections, and it is likely that the mechanism 
for production of conventional charmonium is much more
involved~\cite{Chung:2018lyq, Chung:2022uih}. 
In contrast, $\chi_{c1}(3872)$ production processes could directly probe
perturbative QCD through production of $c \bar c$ in the $^3S_1^{[8]}$ state. 
For example, we can already see that the inclusion of important radiative
corrections including threshold and DGLAP logarithms to all orders in
perturbation theory pioneered in Refs.~\cite{Bodwin:2014gia, Bodwin:2015iua,
Chen:2021hzo, Chung:2024jfk} significantly improves the QCD description of the
nonprompt fraction, compared to strictly fixed-order calculations. 
We anticipate that our knowledge of perturbative $c \bar c$
production rates honed through decades of quarkonium production phenomenology 
can finally be turned into a scalpel to dissect QCD through hadroproduction of 
$\chi_{c1}(3872)$.

\begin{acknowledgments}
We thank Abhishek Mohapatra for useful discussions. 
The work of H.~S.~C. is supported by the 
Basic Science Research Program through the National Research Foundation of
Korea (NRF) funded by the Ministry of Education (Grant No. RS-2023-00248313).
W.~K.~L. is supported by the National Natural Science Foundation of China (NSFC) under Grant No. 12035007 and by the Guangdong Major Project of Basic and Applied Basic Research No. 2020B0301030008.

\end{acknowledgments}

\bibliography{xtetra.bib}

\end{document}